\documentclass[11pt]{article}

\usepackage{amsfonts,amssymb,amsmath,epsfig,epic,color,amsthm}
\usepackage{graphicx}
\usepackage{dcolumn}
\usepackage{bm}

 \usepackage[titletoc, title]{appendix}

\DeclareMathOperator\tr{Tr}
\DeclareMathOperator\diag{diag}

\theoremstyle{plain}

\theoremstyle{definition}

\theoremstyle{remark}

\newcommand{\be}{\begin{equation}}
\newcommand{\ee}{\end{equation}}
\newcommand{\beqs}{\begin{eqnarray}}
\newcommand{\eeqs}{\end{eqnarray}}

\begin{document}

\title{\bf A Singular One-Dimensional Bound State Problem and its Degeneracies}

\author{\centerline {Fatih Erman$^1$, Manuel Gadella$^{2}$, Se\c{c}il Tunal{\i}$^3$, Haydar Uncu$^4$}
\\\and
 {\scriptsize{$^1$
Department of Mathematics, \.{I}zmir Institute of Technology, Urla,
35430, \.{I}zmir, Turkey}}
\\\and
 {\scriptsize{$^2$
Departamento de F\'isica Te\'orica, At\'omica y \'Optica and IMUVA. Universidad de Valladolid,}} \\\and 
 {\scriptsize{Campus Miguel Delibes, Paseo Bel\'en 7, 47011, Valladolid, Spain}}
\\\and
 {\scriptsize{$^3$
Department of Mathematics, \.{I}stanbul Bilgi University, Dolapdere Campus
34440 Beyo\u{g}lu, \.{I}stanbul}, Turkey}
\\\and
 {\scriptsize{$^4$
Department of Physics, Adnan Menderes University, 09100, Ayd{\i}n, Turkey}}
\\
{\scriptsize{E-mail: fatih.erman@gmail.com, manuelgadella1@gmail.com, secil.tunali@bilgi.edu.tr, huncu@adu.edu.tr}}}

\maketitle

\begin{abstract}
We give a brief exposition of the formulation of the  bound state problem for the one-dimensional system of $N$ attractive Dirac delta potentials, as an $N \times N$ matrix eigenvalue problem ($\Phi A =\omega A$). The main aim of this paper is to illustrate that the non-degeneracy theorem in one dimension breaks down for the equidistantly distributed Dirac delta potential, where the matrix $\Phi$ becomes a special form of the circulant matrix.
We then give elementary proof that the ground state is always non-degenerate and the associated wave function may be chosen to be positive by using the Perron-Frobenius theorem. We also prove that removing a single center from the system of $N$ delta centers shifts all the bound state energy levels upward as a simple consequence of the Cauchy interlacing theorem.
\end{abstract}


\section{Introduction}

Dirac delta potentials or point interactions, or sometimes called contact potentials are one of the exactly solvable classes of idealized potentials, and are used as a pedagogical tool to illustrate various physically important phenomena, where the de Broglie wavelength of the particle is much larger than the range of the interaction. They have various applications in almost all areas of physics, see e. g., \cite{Demkov} and \cite{Belloni Robinett}, and references therein. For instance, mutually non-interacting electrons moving in a fixed crystal can be modeled by periodic Dirac delta potentials, known as the Kronig-Penney model \cite{KPmodel}.  Another application is given by the model consisting of two attractive Dirac delta potentials in one dimension. This is used as a very elementary model of the chemical bond for a diatomic ion ($H_{2}^{+}$, for example) and has been discussed in \cite{CohenT, L1}.

The interest for Dirac delta potentials and other
one dimensional point potentials provides us with solvable (or
quasi-solvable) models in quantum mechanics that give insight
for a better understanding of the basic features of the quantum theory.
This makes them suitable for the purpose of teaching the
discipline. In a recent pedagogical review \cite{Belloni Robinett}, several interesting features of one-dimensional Dirac delta potentials have been illustrated and the multiple $\delta$-function potential has been studied in the Fourier space. Moreover, the bound state problem has been formulated in terms of a matrix eigenvalue problem.

In this paper, we first give a brief review of the bound state spectrum of the $N$ Dirac delta potentials in one dimension by converting the time independent Schr\"{o}dinger equation $H \psi = E \psi$ for the bound states to the eigenvalue problem for an $N \times N$ Hermitian matrix.  
This method is rather useful especially when we deal with a large number of centers since the procedure that uses the matching conditions for the wave function at the location of the delta centers become cumbersome for large values of $N$.  Once we formulate the problem as a finite dimensional eigenvalue problem, we show that there are at most $N$ bound states for $N$ centers, using the Feynman-Hellmann theorem (see page 288 in \cite{Griffiths}). One of the main purposes of this paper is to show that this simple one-dimensional toy problem for more than three centers allows us to give an analytical example of the breakdown of the well-known non-degeneracy theorem for one-dimensional bound state problems\cite{LL}. 
This shows that we should not take the non-degeneracy theorem for granted particularly for singular interactions. This was first realized for the so-called one-dimensional Hydrogen atom \cite{Loudon}, where the non-degeneracy theorem breaks down and has been studied for other one-dimensional singular potentials since then \cite{KJSQT, Cohen, Bhattacharyya, Kar, Vincenzo, Dutt}. 
In contrast to the degeneracies that appear in bound states, we give elementary proof that the ground state is non-degenerate and the ground state wave function can be always chosen as real-valued and strictly positive.  In addition, we also show that all the bound state energies for $N$ attractive Dirac delta potentials increase if we remove one center from the system. 
All these results become more transparent using some basic theorems from linear algebra, namely  the Perron-Frobenius theorem, the Cauchy interlacing theorem \cite{Meyer}. Simple proof for these theorems is given in the appendices so as not to interrupt the flow of the presentation. Our presentation is kept simple so that it is also accessible to a wide audience and it has been shown in appendix A that it is consistent with the rigorous approach to the point interactions \cite{Albeverio}.

\section{Bound States for $N$ Dirac Delta Potentials}
\label{Bound States for N Dirac Delta Potentials}

We consider a particle moving in one dimension and interacting with the attractive $N$ Dirac delta potentials located at $a_i$ with strengths $\lambda_i>0$, where $i=1,2, \cdots, N$. The time independent Schr\"{o}dinger equation is then given by
\beqs
-{\hbar^2 \over 2m} {d^2 \psi \over d x^2} - \sum_{i=1}^{N} \lambda_i \delta(x-a_i) \psi (x) = E \psi(x) \;. \label{Schrodinger}
\eeqs
The above equation is actually a formal expression and its exact meaning can only be given by self-adjoint extension theory \cite{Albeverio, BFV, ACP}. Here we follow a more traditional and heuristic approach used in most quantum mechanics textbooks since the results that we obtain is completely consistent with the rigorous approach. As is well-known, the above equation can also be written as 
\beqs H \psi =E \psi \label{hpsi}\eeqs
in the operator form, where  $H={P^2 \over 2m} + V$ and the potential energy operator $V$ for the above particular case in the bra-ket formalism is
\begin{equation}\label{5}
V=- \sum_{i=1}^{N} \lambda_i|a_i\rangle\langle a_i| \,. 
\end{equation}
Here $|a_i\rangle$ is the position eigenket. In the coordinate representation, the action of $V$ on the state vector $|\psi \rangle$ is 
\beqs (V \psi) (x) =
\langle x | V | \psi \rangle =\sum_{i=1}^{N} \lambda_i \delta(x-a_i) \psi(a_i) = \sum_{i=1}^{N} \lambda_i \delta(x-a_i) \psi(x) \;,
\eeqs
where we have used the fact $\delta(x-a_i) \psi(a_i)=\delta(x-a_i) \psi(x)$. This justifies the above formal potential operator (\ref{5}) which corresponds to the Schr\"{o}dinger equation (\ref{Schrodinger}) with multiple Dirac delta potentials. 
Let us absorb the strengths $\lambda_i$'s into bras and kets, i.e., $\sqrt{\lambda_i} |a_i \rangle=|f_i \rangle$ and similarly for bras. In terms of the rescaled bras and kets, the potential operator becomes $V=\sum_{i=1}^{N}|f_i \rangle \langle f_i|$. Substituting this into (\ref{hpsi}) in the coordinate representation, we obtain
\begin{equation}\label{6}
\langle x | {P^2 \over 2m} |\psi \rangle - \sum_{i=1}^{N}  \langle x |f_i\rangle\langle f_i| \psi \rangle = E \langle x | \psi \rangle \;,
\end{equation}
The rescaling is introduced to formulate the bound state problem in terms of an eigenvalue problem of a symmetric matrix, as we will see.  Inserting the
completeness relation $\int {d p \over 2 \pi \hbar} |p \rangle
\langle p|=1$ in front of $|\psi\rangle$ and $|f_i\rangle$, we
obtain the following integral equation, which is actually the
Fourier transformation:
\begin{eqnarray}\label{7}
\int_{-\infty}^{\infty} {d p \over 2 \pi \hbar}\;  e^{{i \over
\hbar} p x} \; \tilde{\psi}(p) \left( {p^2 \over 2m} - E \right) =
\sum_{i=1}^{N} \sqrt{\lambda_i} \; \int_{-\infty}^{\infty} {d p
\over 2 \pi \hbar}\;  e^{{i\over \hbar} p (x-a_i)} \; \phi(a_i) \;
\end{eqnarray}
where $\langle x |p \rangle = e^{{i\over \hbar} p x}$, $\langle p
| \psi \rangle =\tilde{\psi}(p)$, and $\phi(a_i)=\langle f_i | \psi \rangle = \sqrt{\lambda_i}
\psi(a_i)$. Since two functions with the same Fourier transforms
are equal, equation (\ref{7}) implies that:
\begin{eqnarray}
\tilde{\psi}(p)= \sum_{i=1}^{N} \sqrt{\lambda_i} \; { e^{-{i\over \hbar} p a_i} \over {p^2 \over 2m}-E}\; \phi(a_i) \;.\label{psi(p)}
\end{eqnarray}
It is interesting to remark that this solution depends on the
unknown coordinate  wave function at $a_i$ and the energy $E$. If
we use the relation between the coordinate and momentum space wave
function through the Fourier transformation
\begin{eqnarray}
\psi(x)= \int_{-\infty}^{\infty} {d p \over 2 \pi \hbar}\;  e^{{i\over \hbar} p x}  \; \tilde{\psi}(p) \;, \label{Fourier}
\end{eqnarray}
and insert (\ref{psi(p)}) into the above for $x=a_i$, we obtain the following consistency relation
\begin{eqnarray}\label{10}
\psi(a_i)= \sum_{j=1}^{N} \sqrt{\lambda_j} \int_{-\infty}^{\infty} {d p \over 2 \pi \hbar}\; { e^{{i \over \hbar} p (a_i-a_j)} \over {p^2 \over 2m}-E} \; \phi(a_j) \;.
\end{eqnarray}
Multiplying both sides of (\ref{10}) by $\sqrt{\lambda_i}$ and separating the $(j=i)$-th term, we have
\begin{eqnarray}
\left[1- \lambda_i \int_{-\infty}^{\infty}  {d p \over 2 \pi \hbar }  \; \frac{1}{\frac{p^2}{2m}- E} \right] \phi(a_i) -\int_{-\infty}^{\infty}  {d p \over 2 \pi \hbar} \; \sum_{\substack{j=1 \\
j \neq i}}^{N} \sqrt{\lambda_i \lambda_j} \left[\frac{e^{{i\over \hbar} p(a_{i}-a_{j})}
}{\frac{p^2}{2m}-E} \; \right]\phi(a_j)=0 \;. \label{ai aj equation in one}
\end{eqnarray}
This equation can be written as a homogeneous system of linear
equations in matrix  form:
\begin{eqnarray}\label{12}
\sum_{j=1}^{N} \Phi_{ij}(E) \phi(a_j)=0 \;,
\end{eqnarray}
where
\begin{eqnarray}
\Phi_{ij}(E)=
\begin{cases}
\begin{split}
1 - \lambda_i \int_{-\infty}^{\infty}  {d p \over 2 \pi \hbar }  \; \frac{1}{\frac{p^2}{2m}-E}
\end{split}
& \textrm{if $i = j$}\;, \\ \\
\begin{split}
- \sqrt{\lambda_i \lambda_j} \int_{-\infty}^{\infty}  {d p \over 2 \pi \hbar }  \; \frac{e^{{i\over \hbar} p(a_{i}-a_{j})}}{\frac{p^2}{2m}- E}
\end{split}
& \textrm{if $i \neq j$}\;. \label{principle matrix in one}
\end{cases}
\end{eqnarray}
As usual, the matrix elements
are denoted by $\Phi_{ij}(E)$ and the matrix itself by $\Phi$, so
that $\Phi=\{\Phi_{ij}(E)\}$. 
Let us first assume that $E<0$, i.e., $E=-|E|$, so that
there is no real pole in the denominators of the integrands. Let us now
consider the integral in the off-diagonal part.
The function under the
integral sign has simple poles located at the points ($p=\pm i
\sqrt{2m|E|}$) in the complex $p$-plane. In order to calculate this integral by the residue
method, we have to take into account separately the situations $a_i<a_j$ and
$a_i>a_j$. 
We note that only the pole with sign plus (minus) lies inside the
contour of integration for $a_i>a_j$ ($a_j>a_i$). Due to the exponential function, the integral
over the semicircle vanishes as its radius goes to infinite
\cite{ASH}. Then, the value of the integral is obtained
multiplying by $2\pi i$ the residue at that point:
\begin{eqnarray}
\label{14}
\int_{-\infty}^{\infty}  {d p \over 2 \pi \hbar }  \; \frac{e^{{i
\over \hbar}p(a_{i}-a_{j})}}{\frac{p^2}{2m}- E} =
\begin{cases}
\begin{split}
 {m \over \hbar
\sqrt{2m|E|}} \exp \left(-\sqrt{2m |E|} (a_i-a_j)/\hbar \right)\;,
\end{split}
& \textrm{if $a_i >a_ j$}\;, \\ \\
\begin{split}
 {m \over \hbar
\sqrt{2m|E|}} \exp \left(-\sqrt{2m |E|} (a_j-a_i)/\hbar \right)\;,
\end{split}
& \textrm{if $a_i <a_j$}\;.
\end{cases}
\end{eqnarray}
The diagonal part of the matrix $\Phi$ can be evaluated similarly, so equation (\ref{principle matrix in one}) becomes:
\begin{eqnarray}\label{15}
\Phi_{ij}(E)=
\begin{cases}
\begin{split}
1 - {m \lambda_i \over \hbar \sqrt{2m|E|}}
\end{split}
& \textrm{if $i = j$}\;, \\ \\
\begin{split}
-{m \sqrt{\lambda_i \lambda_j} \over \hbar \sqrt{2m|E|}} \; \exp
\left(-\sqrt{2m |E|} |a_i-a_j|/\hbar \right)
\end{split}
& \textrm{if $i \neq j$}\;.
\end{cases}
\end{eqnarray}
Equation (\ref{12}) has only non-trivial solutions if $\det
\Phi (E)=0$.  Therefore, the bound state problem is solved once we
find the solution to the transcendental equation $\det \Phi
(E)=0$. After that, we can find the bound state wave
functions in the coordinate representation
through (\ref{Fourier}). 
Suppose that the bound state energy, say $E_B$, is the root of $\det \Phi
(E)=0$, and we find $\phi_B(a_j)=\sqrt{\lambda_j} \psi_B(a_j)$ from Eq.(\ref{12}) associated with $E_B$. Then, the bound state wave function at $a_i$ is 
\begin{eqnarray}
\psi_B(a_i)= {1 \over \sqrt{\lambda_i}} \; \phi_B(a_i) \;. \label{bswavefunctionatpoint}
\end{eqnarray}  
Taking into account the above considerations, we use (\ref{bswavefunctionatpoint}) into the bound state wave function in momentum space (\ref{psi(p)}) so as to obtain
\begin{eqnarray}
\tilde{\psi}_B(p)=\sum_{i=1}^{N} \sqrt{\lambda_i} \; { e^{-{i\over \hbar} p a_i} \over {p^2 \over 2m}-E_B}\; \phi_B(a_i) \;. \label{bswavefunctioninmomentumspace}
\end{eqnarray}
Then, the bound state wave function in the coordinate space can be found by just taking the inverse Fourier transform of the above momentum space wave function
\begin{eqnarray}
\psi_B(x)= \sum_{i=1}^{N} \lambda_i \; \phi_B(a_i)\; \sqrt{m/2} \; \;  {e^{-{\sqrt{2m |E_B|} \over \hbar} |x-a_i|} \over \hbar \sqrt{|E_B|}} \;,
\end{eqnarray}
where $\phi_B(a_i)$ is defined by $\sum_{j=1}^{N} \Phi_{ij}(-|E_B|) \; \phi_B(a_j)=0$.
Suppose now that $E>0$. In this case, we have to find the wave function and contour integrals of the form 
\begin{equation}
\int_{-\infty}^{\infty} {d p \over 2 \pi \hbar} {e^{{i \over \hbar} p(a_i-a_j)} \over {p^2 \over 2m}-E} \;,
\end{equation} 
whose poles are now located at $p=\pm \sqrt{2m E}$ on the real axis, and there are four different choices of contours, each of which gives different result \cite{DenneryKrzywicki}. 
It is easy to see that the wave function  
becomes now the linear combination of the complex exponentials 
\begin{equation}
e^{\pm i  \sqrt{2 m E/\hbar}|x-a_i|} \;.
\end{equation}
Such a function cannot be square integrable unless it is identically zero . Therefore, there is no bound state for $E>0$. A similar analysis can be done for $E=0$, where the wave function becomes divergent over the whole real axis. Therefore, we conclude that $E$ must be negative for bound states. From the physical point of view, the bound state energies are expected to be less than the values of the potential at asymptotes. For this reason, the bound state energies for finitely many Dirac delta potentials are negative.

For a single center located at $x=0$ with coupling constant $\lambda$, the matrix $\Phi$ is just a $1 \times 1$ matrix, i.e., a single function: $\Phi(E)=1 - {m \lambda \over \hbar \sqrt{2m|E|}}$. Now, the condition $\det \Phi(E)=0$ means that $\Phi(E)=0$,  so that the bound state energy is $E_B= -{m \lambda^2 \over 2 \hbar^2}$ for a single center \cite{Griffiths}.  After having found the bound state energy, we can find the bound state wave function. For $N=1$, $\phi_B(a_i)$ is some constant, say $C$. Then, the bound state wave function in momentum space becomes
\begin{eqnarray}
\tilde{\psi}_B(p)=\sqrt{\lambda} \; {1 \over {p^2 \over 2m}-E_B} \; C \;. \label{bswavefunctioninmomentumspaceN1}
\end{eqnarray}
The constant $C$ can be determined from the normalization constant:
\begin{eqnarray}
\lambda \; |C|^2 (2m)^2 \int_{-\infty}^{\infty} {d p \over 2 \pi \hbar} {1 \over (p^2+{m^2 \lambda^2 \over \hbar^2})^2} =1 \;. \label{normalizationcond}
\end{eqnarray} 
The integral in Eq. \eqref{normalizationcond} can also be evaluated using the residue theorem. However, in this case the residues are at $p=\pm i {m \lambda \over \hbar}$ and of order two. Taking the integral, we find $C={\sqrt{m \lambda} \over \hbar}$. 
Now we can find the wave function associated with this bound state in the coordinate space by taking its Fourier transform. We perform the integration exactly as we did in (\ref{14}), and obtain \cite{Griffiths}
\begin{eqnarray}
\psi_B(x) = {\sqrt{m \lambda} \over \hbar} \; e^{-{m \lambda \over \hbar^2} |x|} \;.
\end{eqnarray}

Let us first consider the special case of two centers, namely twin attractive ($\lambda_{1}=\lambda_{2}=\lambda$) Dirac $\delta$ potentials located at $a_{1}=0$ and
$a_{2}=a$. Then, the expression $\det \Phi (E)=0$ yields to the
following transcendental equation:
\begin{eqnarray}\label{16}
e^{-\frac{a\sqrt{2m|E|}}{\hbar}}=
\pm\left( {\hbar \sqrt{2 m |E|} \over m \lambda}-1\right) \;.
\end{eqnarray}
For convenience, we define $\kappa \equiv {\sqrt{2m|E|} \over \hbar}$. Suppose that $\kappa_+(\kappa_-)$ corresponds to the solution of Eq. (\ref{16}) with the positive (negative) sign 
in front of the parenthesis, i.e.,
\begin{eqnarray}\label{17}
e^{-a\kappa_{+}} &=& \frac{\hbar^{2}\kappa_{+}}{m\lambda}-1 \;,
\end{eqnarray}
or
\begin{eqnarray}
e^{-a\kappa_{-}} &=& 1-\frac{\hbar^{2}\kappa_{-}}{m\lambda} \,.
\label{18}
\end{eqnarray} 
The bound state energies correspond to non-zero solutions for $\kappa_\pm$ of the above equations (\ref{17}) and (\ref{18}). The first transcendental equation (\ref{17})  always has
one real root, which implies the presence of at least one bound state. This is clear from the following considerations: 
 the left hand side of (\ref{17}) is a monotonically
decreasing function, which goes to zero asymptotically, while the
right hand side is a monotonically increasing function without any
asymptote.
However, the second transcendental equation (\ref{18}) may or may
not have a real positive solution.
One real root of Eq. (\ref{18}) is expected for
$\kappa_- =0$. However, this cannot correspond to a bound state. In order to obtain a non trivial root, we must impose the
condition that the slope of the right hand side of (\ref{18}) must
be smaller than the slope of the left hand side in absolute value. 
\begin{eqnarray}\label{19}
\bigg|{d \over d \kappa} \left(1-{\hbar^2 \kappa \over m \lambda}\right)
\bigg|_{\kappa=0} \bigg| & < & \bigg|{d \over d \kappa} \left(e^{-\kappa a}\right)
\bigg|_{\kappa=0} \bigg|\,.
\end{eqnarray}
This means that the distance between the centers must be greater than some critical value for two bound states:
\begin{equation} \label{19a}
a>  {\hbar^2 \over m \lambda} \;.
\end{equation}
Hence, we conclude that there are at most two bound states for
attractive twin Dirac delta potentials. The first one
appears unconditionally so that it corresponds  to the ground
state. On the other hand,  the second bound state
appears only if $a$ is sufficiently large ($ {\hbar^2 \over  m a \lambda} < 1$). This
corresponds to the excited state of the system.

Actually, the explicit solutions to Eq.(\ref{17}) and Eq.(\ref{18}) can be easily found and then the bound state energies are
\begin{eqnarray}
E_+ & = & - \left( {\lambda \over 2} + {1 \over a} \; W \left[{a \lambda \over 2} e^{-{a \lambda \over 2}}\right] \right)^2 \;, \cr 
E_- & = & - \left( {\lambda \over 2} + {1 \over a} \; W \left[-{a \lambda \over 2} e^{-{a \lambda \over 2}}\right] \right)^2 \;. \label{lambertW}
\end{eqnarray}
where $W$ is the Lambert $W$ function  \cite{Corless}, defined as the solution of the transcendental equation $y \; e^y =z$, i.e., $y=W[z]$. The above explicit solutions given in terms of Lambert $W$ function have been known in the literature, see for instance \cite{SecilTunali}
and the recent work \cite{Sacchetti}, where the non-linear generalization of the problem has been discussed.

\section{Bound States as a Finite Dimensional Eigenvalue Problem}

In order to study location and properties of bound states  more systematically, we consider the equation (\ref{12}) as the particular case of an eigenvalue problem for the matrix $\Phi$:
\begin{equation} \label{23}
\Phi(E) \, A(E) = \omega(E)\,  A(E) \;,
\end{equation}
where $\omega$ is any of  the eigenvalues of the matrix $\Phi$. Then, the zeros of the eigenvalues of $\Phi$ are just the bound state energies. In other words, the roots of the equation
\begin{equation}
\omega(E)=0 \label{zeroeigenvalue}
\end{equation}
give the bound state energies. Hence, the eigenvalues of the {\it linear}
differential equation $H\psi(x)=E\psi(x)$ are obtained through a
{\it non-linear} transcendental algebraic problem, $\omega(E)=0$.

Let us consider the $N=2$ case. For twin centers, located at $a_1=0, a_2=a$, the eigenvalues can be explicitly calculated:
\begin{eqnarray} \label{eigenvaluesformulaN2}
\omega_1 & = & 1+ {m \lambda \over \hbar \sqrt{2m|E|}} \left( -1-e^{-{1 \over \hbar} \sqrt{2m |E|}a} \right)\cr \omega_2 & =& 1+ {m \lambda \over \hbar \sqrt{2m|E|}} \left( -1+e^{-{1 \over \hbar} \sqrt{2m |E|}a} \right)  
\end{eqnarray}

\begin{figure}[h!]
\centering
\begin{minipage}{5cm}
\includegraphics[scale=0.5]{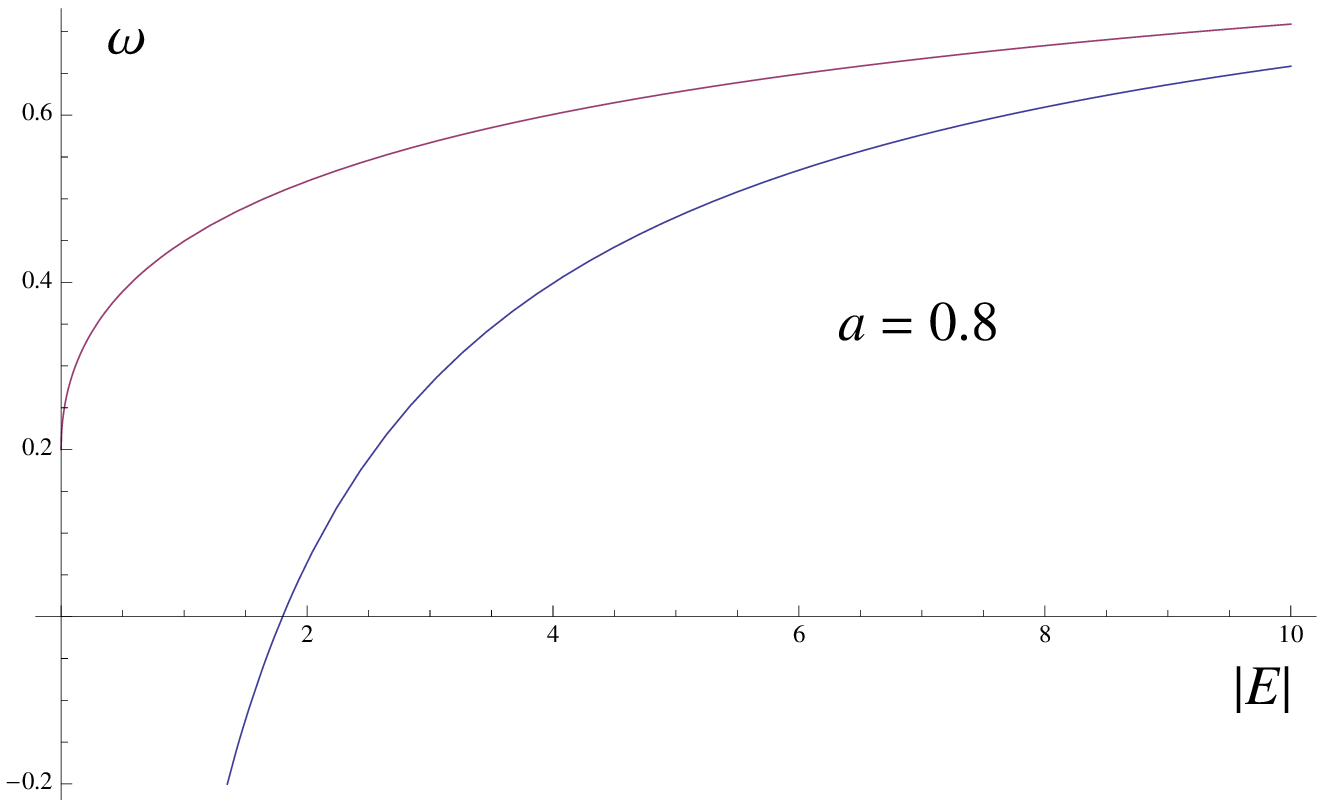}
\end{minipage}
\qquad \qquad \qquad 
\begin{minipage}{5cm}
\includegraphics[scale=0.5]{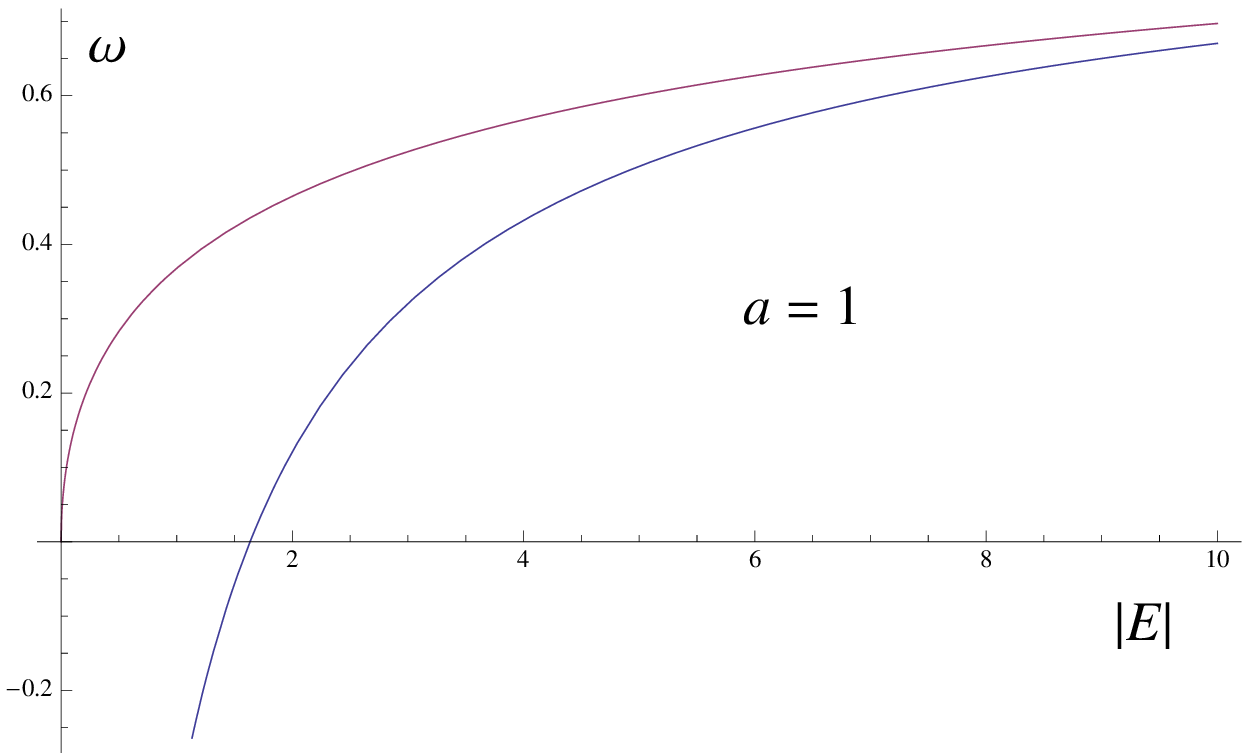}
\end{minipage} \\
\begin{minipage}{5cm}
\includegraphics[scale=0.5]{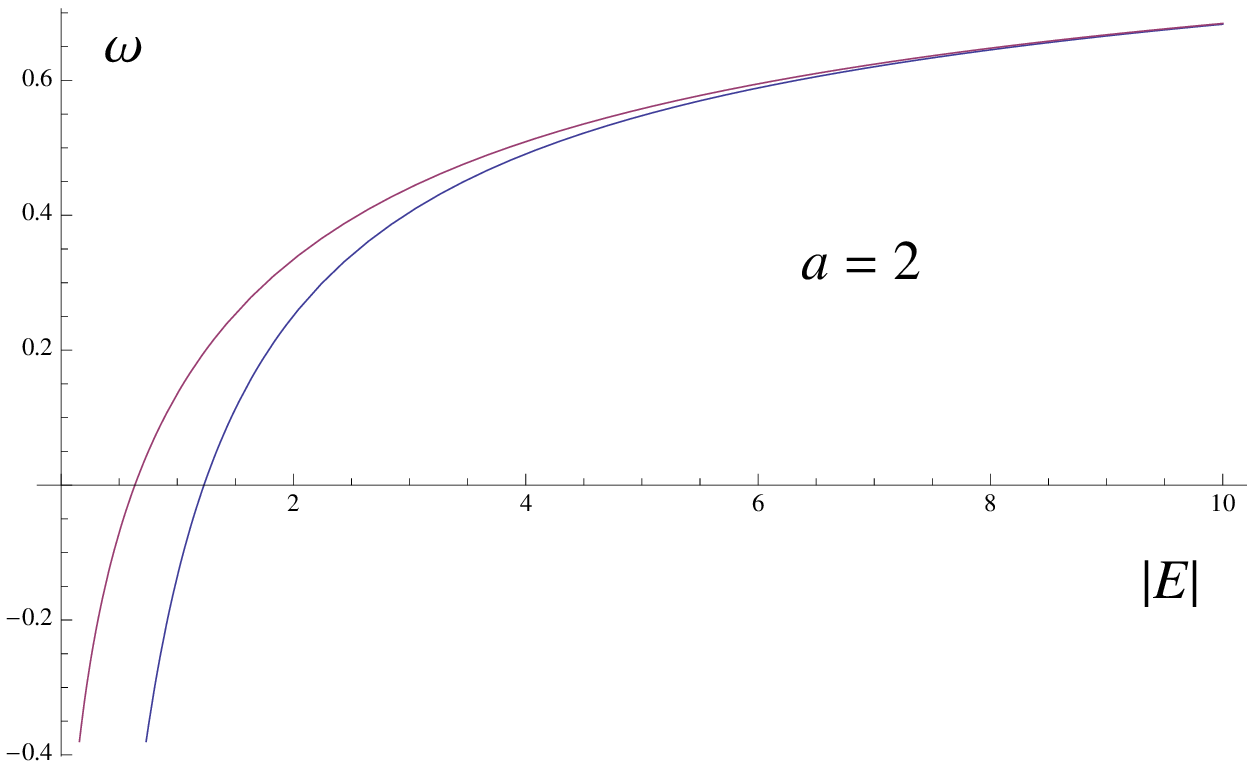}
\end{minipage}
\qquad \qquad \qquad
\begin{minipage}{5cm}
\includegraphics[scale=0.5]{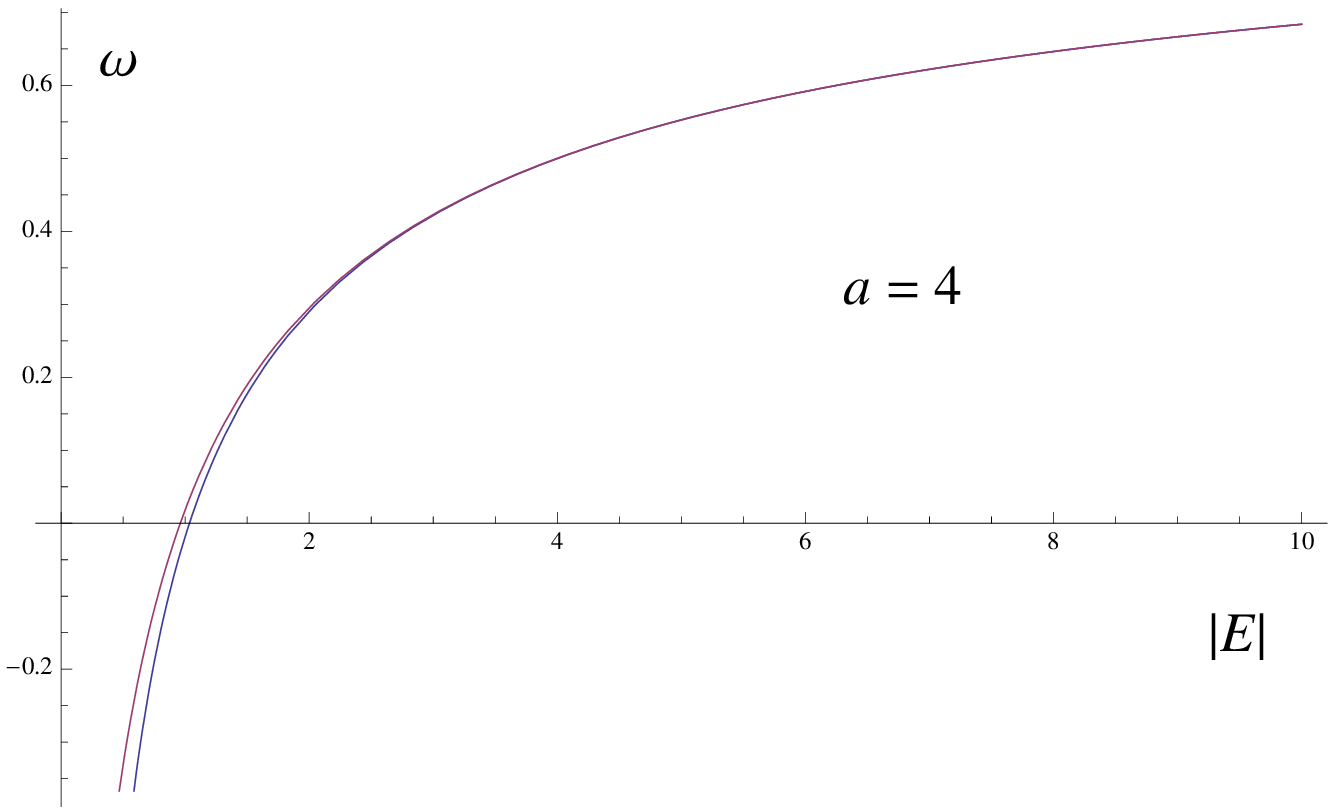}
\end{minipage}
\caption{The flow of the eigenvalues of the matrix $\Phi$ as a function of $|E|$ for different values of  $a$. Here $\lambda=2$ and $\hbar=2m=1$.} \label{eigenvaluesN2}
\end{figure}
As shown in above Fig. \ref{eigenvaluesN2}, there are always two eigenvalues of the matrix $\Phi$. However, for $\lambda=2$ with $\hbar=2m=1$, there are two bound states only  if the distance between the centers is greater than the critical value $a=1$. Otherwise there is only one bound state, which is consistent with the result given in the previous part. 
When the centers are sufficiently close to each other,  one of the bound states seems to disappear, since the zeros of the first eigenvalue seems to move to the negative real axis.

We also observe from Fig. \ref{eigenvaluesN2} that the bound state energies come closer and closer as the distance between them increases. This is not surprising since the eigenvalues (\ref{eigenvaluesformulaN2}) converge to $1- {m \lambda \over \hbar \sqrt{2m|E|}} $ as $a \rightarrow \infty$ so that zeros of these degenerate eigenvalues lead to degenerate bound states in the limiting case.

Now, we shall show why our method is much easier to investigate the bound state spectrum as we
increase the number of Dirac delta potentials. The number of bound
states is an important characteristic of any system. There are
several ways to determine it for some regular potentials
\cite{Manoukian}. It is noteworthy that we can determine the maximum
number of bound states of this system from
the behavior of the eigenvalues of the matrix $\Phi(E)$ through the Feynman-Hellmann theorem \cite{feynman, hellmann}.  

To find the behavior of the eigenvalues as a function of $E$, let us first take the derivative  of $\Phi_{ij}(E)$  with
respect to $E$. We may interchange this derivative and the
integral sign in (\ref{15}), since all the matrix elements
$\Phi_{ij}$ are analytic functions on the half plane $\Re{(E)}<0$.  Hence, we obtain
\begin{eqnarray}\label{24}
{d \Phi_{ij} \over d E} = - \sqrt{\lambda_i \lambda_j} \int_{-\infty}^{\infty}  {d p \over 2 \pi \hbar }
\; \frac{e^{{i\over \hbar} p(a_{i}-a_{j})}}{(\frac{p^2}{2m}- E)^2} \;.
\end{eqnarray}
Now, let us make use of the Feynman-Hellmann theorem, which states that
\begin{eqnarray}\label{25}
{d \omega(E)\over d E} = \langle A^k |  {d \Phi_{ij} \over d E} | A^k \rangle \;,
\end{eqnarray}
where $A^k$ is a given normalized eigenvector for $\omega(E)$. In other words, the Feynman-Hellmann theorem states that 
the derivative
of the eigenvalue of a parameter dependent Hermitian matrix 
is equal to the expectation value of the derivative of
the matrix with respect to its normalized eigenvector.
The Feynman-Hellmann theorem can be generalized for the degenerate states \cite{Vatsya} but it does not change our conclusion that we will draw. Thus, we have
\begin{eqnarray}\label{26}
{d \omega(E) \over d E} &=& - \sqrt{\lambda_i \lambda_j} \sum_{i,j=1}^{N} (A^{k}_{i})^{*}  \int_{-\infty}^{\infty}  {d p \over 2 \pi \hbar }  \; \frac{e^{{i\over \hbar} p(a_{i}-a_{j})}}{(\frac{p^2}{2m}- E)^2} \; A^{k}_{j} \cr & = & - \int_{-\infty}^{\infty}  {d p \over 2 \pi \hbar }  \; \frac{1}{(\frac{p^2}{2m}- E)^2} \; \left| \sum_{i=1}^{N} e^{-{i \over \hbar} p a_{i}} \; \sqrt{\lambda_i} \; A^{k}_{i} \right|^2 < 0\,.
\end{eqnarray}
For $E=-|E|$, ${d \omega(E) \over d |E|} >0$. 
Since there are at most $N$ distinct eigenvalues of the $N \times N$ matrix $\Phi$ and these eigenvalues are
monotonically increasing functions of $|E|$, there must be at most
$N$ bound states. This conclusion would have been rather difficult to arrive
just by following the standard method, in which the properties of
the bound states are just determined by matching conditions at the
locations of the delta centers.

\section{Degeneracies in the Bound States for Periodically Distributed Centers}

Let us consider $N$ Dirac delta potentials located equidistantly, i.e., $a_0=0, a_1=a, a_2=2a, \ldots, a_N=(N-1) a$ and $\lambda_1=\ldots=\lambda_N=\lambda$. Then, the matrix $\Phi$ given in Eq. (\ref{15}) takes the following form
\begin{eqnarray}
\left(
\begin{array}{ccccc}
 c_0 & c_1 & \cdots  & c_{N-2} & c_{N-1} \\
 c_{N-1} & c_0 & c_1 & \cdots  & c_{N-2} \\
 c_{N-2} & c_{N-1} & \ddots & \ddots & \vdots  \\
 \vdots  & \vdots  & \ddots & \ddots & c_1 \\
 c_1 & c_2 & \cdots  & c_{N-1} & c_0
\end{array}
\right)_{N\times N} \label{circulantmatrix}
\end{eqnarray}
where $c_0=1 - {m \lambda \over \hbar \sqrt{2m|E|}}$ and 
\beqs c_j=c_{N-j}=-{m \lambda \over \hbar \sqrt{2m|E|}} \; \exp
\left(-\sqrt{2m |E|} j a/\hbar \right) \eeqs 
for all $j=1,\ldots, N-1$. The form of the matrix above (\ref{circulantmatrix}) is usually known as the circulant matrix. By using the Fourier matrix, it can be diagonalized and its eigenvalues can be found easily \cite{Meyer}. However, showing this is the beyond the scope of the main aim of this paper. Nevertheless, it is a simple exercise to show that  
\begin{eqnarray}
\left(
\begin{array}{ccccc}
 c_0 & c_1 & \cdots  & c_{N-2} & c_{N-1} \\
 c_{N-1} & c_0 & c_1 & \cdots  & c_{N-2} \\
c_{N-2} & c_{N-1} & \ddots & \ddots & \vdots  \\
 \vdots  & \vdots  & \ddots & \ddots & c_1 \\
 c_1 & c_2 & \cdots  & c_{N-1} & c_0
\end{array}
\right) \left( \begin{array}{c}  1 \\ \zeta^l \\ \zeta^{2l} \\ \vdots \\ \zeta^{(N-1)l} \end{array}\right) = \lambda_l \; \left( \begin{array}{c}  1 \\ \zeta^l \\ \zeta^{2l} \\ \vdots \\ \zeta^{(N-1)l} \end{array}\right) \;, \label{eigenvaluesofcirculantPhimatrix}
\end{eqnarray}
where $\zeta$ is the $N$ th root of the unity, i.e., $\zeta=e^{2 \pi i/N}$ and $j=0,1, \ldots, N-1$ and $\lambda_l$'s are the eigenvalues of the matrix $\Phi$, given by 
\begin{eqnarray}
\omega_j =\sum_{k=0}^{N-1} c_k \; \zeta^{j k} \;.
\end{eqnarray}
Note that the above formula is reduced to   (\ref{eigenvaluesformulaN2}) for $N=2$. We realize that for $N \geq 3$, the eigenvalues are degenerate 
\begin{equation}
\omega_j=\omega_{N-j}
\end{equation}
for all $j=1,\ldots, N-1$ since 
\begin{eqnarray}
\omega_j = c_0 + \sum_{k=1}^{N-1} c_k \; \zeta^{j k} = c_0 + \sum_{k=1}^{N-1} c_{N-k} \; \zeta^{j k} = c_0 + \sum_{l=1}^{N-1} c_{l} \; \zeta^{j (N-l)}=c_0 + \sum_{l=1}^{N-1} c_{l} \; \zeta^{(N-j)l} =\omega_{N-j} \;,
\end{eqnarray}
where we have used $\zeta^N=1$ and $j(N-l)=(N-j)l \mod N$  \cite{DJW}. Note that $\omega_0$ and $\omega_N$ cannot be degenerate. Since the matrix $\Phi$ is Hermitian, its algebraic multiplicity is equal to its geometric multiplicity \footnote{The algebraic multiplicity of an eigenvalue $\lambda$ is the number times it is repeated as a root of the characteristic polynomial, whereas the geometric multiplicity of $\lambda$ is the maximal number of linearly independent eigenvectors associated with $\lambda$.}, the eigenvectors $A_j$ associated with the degenerate eigenvalues $\omega_j$ span the degenerate space. The functions $\omega_j(E)$ are monotonic functions of $|E|$, so there exists a one-to-one relation between $\omega_j(E)$ and its zeros. From the monotonic behavior of the eigenvalues $\omega_j(E)$ and the explicit relation between the bound state wave function  and the eigenvectors $A_j$, we conclude that the bound state energies are degenerate and the dimension of the degeneracy subspace of $\Phi$ is equal to the dimension of the degeneracy subspace of the bound state wave functions.  
This is contrary to the common belief that there is no degeneracy in one-dimensional bound state problems \cite{LL}.

In order to understand why the non-degeneracy theorem breaks down, we first recall the standard proof of the non-degeneracy theorem for one-dimensional bound state problem of a generic potential $V$. Suppose that there are two bound states $\psi_1$ and $\psi_2$ associated with the same energy $E$. Then, it is easy to show \cite{LL} that the Wronskian of these solutions must be equal to a constant $C$, i.e.,
\begin{equation}
W=\psi_2 { d \psi_1 \over d x} - \psi_1 { d \psi_2 \over d x} = C \;,
\end{equation} 
for all $x$. For bound states, we expect that $\psi \rightarrow 0$ as $x \rightarrow \pm \infty$. As a consequence of this, one can conclude that $C=0$ as long as there is no blow up in the derivatives (this is one argument, where the nondegeneracy theorem breaks down for potentials, see \cite{Kar} ). This implies that
\begin{equation}
\psi_2 { d \psi_1 \over d x} = \psi_1 { d \psi_2 \over d x} \;.
\end{equation}
At points, where $\psi_1$ and $\psi_2$ are nonzero, the division to $\psi_1 \psi_2$ is possible so that we have a separable differential equation  
\begin{equation}
{\psi_{1}' \over \psi_1} ={\psi_{2}' \over \psi_2} \;.
\end{equation}
Hence, the solution to this is given by 
$$\psi_1=c \; \psi_2 \;.$$ 
This contradicts with the initial assumption, which proves that there can not be degeneracy in the bound states of one-dimensional systems. This is the well-known standard proof of the nondegeneracy theorem. However, the above solution is not necessarily true at the points where $\psi_1 \psi_2 =0$. Hence, the above solution is valid only between the adjacent zeros of the $\psi_1 \psi_2$. As a result, $c$ is constant in each region where $\psi_1 \psi_2 \neq 0$ but this does not guarantee that the values of $c$ are the same for all regions. We could have multiple constant $c's$, jumping to each other at the location of zeros of $\psi_1 \psi_2$. As Loudon \cite{Loudon} pointed out  that a discontinuous $c$ implies that $\psi_1$ and $\psi_2$ cannot both have a continuous, finite derivative at their zeros. Since $\psi_1$ and $\psi_2$ satisfy the Schr\"{o}dinger equation with the same eigenvalue, one can heuristically expect that the potential $V$ must have a singularity at the zeros of $\psi_1 \psi_2$. Therefore, the nondegeneracy theorem does not have to be valid for potentials having singularities and there exist one-dimensional singular potentials (e.g., one dimensional Hydrogen atom \cite{Loudon}), where degeneracies may occur.

Let us now explicitly show that the above argument is indeed the case for our problem. The explicit form of the wave function associated with bound state energy $E_B=-|E_B|$ can easily be computed by substituting the solution Eq.(\ref{bswavefunctioninmomentumspace}) for $\lambda_1=\lambda_2=\cdots=\lambda_N=\lambda$ and the centers of delta potentials are located equidistantly into Eq.(\ref{Fourier}) in the manuscript, so that we have
\begin{eqnarray}
\psi_B(x) & = & \lambda \int_{-\infty}^{\infty} {d p \over 2 \pi \hbar} e^{{i \over \hbar} p x} \sum_{i=1}^{N} {e^{-{i \over \hbar} p a_i} \over {p^2 \over 2m} + |E_B|} \; \phi_B(a_i) \cr & = & \lambda \sum_{i=1}^{N} \phi_B(a_i) \sqrt{m/2} \;\; {e^{-{\sqrt{2m |E_B|} \over \hbar} |x-a_i|} \over \hbar \sqrt{|E_B|}} \;.
\end{eqnarray}
It is important to notice that $\phi_B$ is the eigenvector of the matrix $\Phi(E)$ associated with its zero eigenvalues (see Eqs. (\ref{23}), (\ref{zeroeigenvalue}),  and Eq. (\ref{12})). As explicitly shown in Eq. (\ref{eigenvaluesofcirculantPhimatrix}), we have
\begin{eqnarray}
\phi_B= \left( \begin{array}{c}  1 \\ \zeta^l \\ \zeta^{2l} \\ \vdots \\ \zeta^{(N-1)l} \end{array}\right) \;,
\end{eqnarray}
where $\zeta$ is the $N$th root of unity and $j=0,1,\cdots, N-1$. As a consequence of this, we obtain 
\begin{eqnarray}
\psi_B(a_i) &=&  { \lambda \sqrt{m/2} \over \hbar \sqrt{|E_B|}} \sum_{i=1}^{N} \phi_B(a_i) \cr &=&  { \lambda \sqrt{m/2} \over \hbar \sqrt{|E_B|}}  \; (1+ \zeta +\zeta^2 + \cdots + \zeta^{(N-1)}) \cr & = & 0 \;, 
\end{eqnarray}
due to the fact that $(1+ \zeta +\zeta^2 + \cdots + \zeta^{(N-1)})=0$ \cite{BrownChurchill}. This shows that the bound state wave function vanishes at the location where the potential has a singularity, namely at the location of the Dirac delta centers. This is exactly the situation where the non-degeneracy theorem may break down as discussed above.
Hence, we explicitly show the reason why non-degeneracy theorem breaks down for our problem when the Dirac delta centers are located equidistantly.

Actually, it has been demonstrated recently that the one-dimensional Hydrogen atom model may not be indeed a counter example of the non-degeneracy theorem when it is investigated by a more rigorous approach, namely the self-adjoint extension theory \cite{PalmaRaff}. 
Although the method we have used in this paper is heuristic rather than a rigorous analysis using the self-adjoint extension theory, where one has to deal with the technicalities of the domain of the unbounded operators, it is completely consistent with the self-adjoint extension treatment of the Dirac delta potentials given in \cite{Albeverio}, as shown in appendix A.

\section{Non-degeneracy of the Ground State}

For a generic distribution of centers, it is not obvious whether the ground state is non-degenerate or not. Here, we shall show that this is indeed the case by using the Perron-Frobenius theorem (see page 661 in \cite{Meyer}) for symmetric matrices. Actually, the proof of the non-degeneracy of the ground states for some class of potentials has been discussed in \cite{reedsimonv4} and the non-degeneracy of the ground state for point interactions has been proved using the positivity preserving semi-groups generated by $\Phi$ and Beurling-Deny conditions in \cite{Albeverio}. Here we give a more elementary proof, which was also used in the two and three dimensional version of the model in \cite{ErmanTurgut}.

Let us first recall the Perron-Frobenius theorem:

Let $A=(a_{ij})$ be an $N\times N$ symmetric matrix with elements $a_{ij}>0$ and let $\lambda$ be the largest eigenvalue. There follows that:
\begin{enumerate}
  \item $\lambda>0$.
  \item There exists a corresponding eigenvector $(x_{j})$ with every component $x_{j}>0$.
  \item $\lambda$ is non-degenerate.
  \item If $\mu$ is any other eigenvalue, $\lambda>|\mu|$.
\end{enumerate}
In order to make our presentation self-contained, an elementary proof (just using the basic knowledge of linear algebra) of this theorem is given in appendix B \cite{Ninio}.

\begin{figure}[h!] \centering
\includegraphics[scale=0.8]{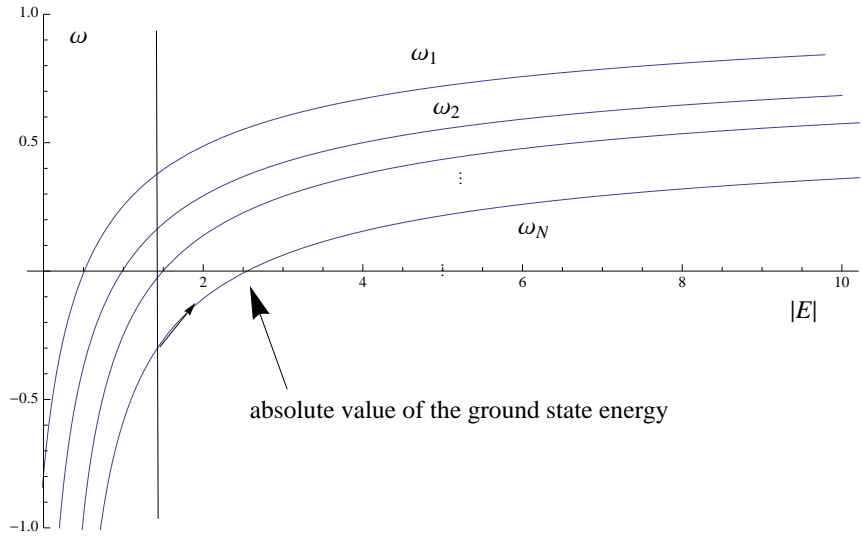}
\caption{Flow of the eigenvalues of $\Phi$} \label{floweigenvaluesfig} 
\end{figure}

Since the matrix $\Phi$ given in (\ref{15}) is symmetric but not positive, we cannot directly apply the Perron-Frobenius theorem. Nevertheless, we can make $\Phi$ positive in such a way that the spectrum of the problem is invariant. One simple way to achieve this is to subtract from $\Phi$ a diagonal matrix whose elements coincide with the maximum of the diagonal elements of $\Phi$ and, then, reversing the
overall sign:
\begin{eqnarray}
\Phi'(E) := -\left[\Phi(E)- (1+ \varepsilon)  
\max_{E_{gr} \leq E < \infty} \;  \diag \left( \Phi_{11} (E), \ldots, \Phi_{NN} (E) \right)   \right] > 0 
\;, 
\end{eqnarray}
where $\varepsilon$ is arbitrarily small positive number. Let $E_{gr}$ be the ground state energy. Since $\Phi_{ii}(E)$ is a decreasing function of $E$, $\max_E \; \Phi_{ii}(E) = \Phi_{ii}(E_{gr})$.  Let us simplify this further by replacing $\Phi_{ii}(E_{gr})$ with $\max_{i} \Phi_{ii}(E_{gr})=:\Phi(E_{gr}, \lambda_{min})$, where $\lambda_{min} :=\min_i \lambda_i$. Here, we have used the fact that $\Phi_{ii}$ is a decreasing function of $\lambda_i$.  Then, we define 
\beqs \Phi''(E):= -\left[\Phi(E)- (1+ \varepsilon) 
\;  I \; \Phi (E_{gr},\lambda_{min})  \right]
\;, \eeqs
where $I$ is the identity matrix. Adding a diagonal matrix to $\Phi$ does not change its eigenvectors whereas its eigenvalues are shifted by a constant amount. However, this is equivalent to an overall translation in the bound state spectrum, which is physically unobservable. 
Hence, the transformed positive matrix $\Phi''$ and $\Phi$ have common eigenvectors and this guarantees that there exist a strictly positive
eigenvector $A$ for  $\Phi''$ and
\beqs \Phi''(E) A(E)= - \left[ \omega(E)- (1+ \varepsilon) 
\;  I \;  \Phi(E_{gr}, \lambda_{min}) \right] \; A(E) = \omega''(E) \; A(E) 
\;. \eeqs
The minimum eigenvalue of $\Phi$ corresponds to the maximum eigenvalue of $\Phi''$. 
For a given $E$, there exists a strictly negative  non-degenerate minimum eigenvalue of $\Phi$, say $\omega^{\min}(E)$ as a consequence of the Perron-Frobenius theorem.  Since we are looking
for the zeros of the eigenvalues $\omega(E)$, $\omega^{min}$ goes to zero at the ground state energy $E_{gr}=-|E_{gr}|$, as can be easily seen in Fig. \ref{floweigenvaluesfig}. In other words, we must have
\begin{equation}
\omega^{min}(E_{gr})=0 \;.
\end{equation}
Then, from the remaining part of the Perron- Frobenius theorem, we conclude that there exists a corresponding positive
eigenvector $A_{i}(E_{gr})$ associated with the non-degenerate minimum eigenvalue $\omega^{min}(E_{gr})$. Using
\begin{eqnarray}
\psi_{gr} (x) & = & \sum_{i=1}^{N} \sqrt{\lambda_i} \int_{-\infty}^{\infty} {d p \over 2 \pi \hbar}\;   \; \left(  { e^{{i\over \hbar} p (x-a_i)}  \over {p^2 \over 2m} +|E_{gr}|}  \right) \; A_i(E_{gr}) \cr & = & \sum_{i=1}^{N} \sqrt{\lambda_i} \;  A_i(E_{gr}) \; \left(  {m \over \hbar \sqrt{2m|E_{gr}|}} \; \exp
\left(-\sqrt{2m |E_{gr}|} |x-a_i|/\hbar \right) \right) \;,
\end{eqnarray}
we also conclude that $\psi_{gr}(x)$ is positive so it has no node.

Another important result about the bound states is that the ground state energy increases when we remove one of the centers from the system. This can be seen from the Cauchy interlacing theorem and the monotonic behavior of the eigenvalues $\omega$ of the matrix $\Phi$. The Cauchy interlacing theorem states the following (page 552 in \cite{Meyer}):

Let $A$ be a Hermitian matrix of order $N$ and let $B$ be a principal submatrix of $A$ of order $N-1$. Let us list  the eigenvalues of $A$ in decreasing order as $\lambda_{N} \leq \lambda_{N-1} \leq \cdots \leq \lambda_{2} \leq \lambda_1 $ and the same with the eigenvalues of $B$ as $\mu_{N} \leq \mu_{N-1} \leq \cdots \leq \mu_{2} $. Then, we have
\begin{eqnarray}
\lambda_{N} \leq \mu_N \leq \lambda_{N-1} \leq \mu_{N-1} \leq \cdots \leq \lambda_{2} \leq \label{cauchyinterlacingtheorem}
\end{eqnarray}
A simple proof of this theorem using the intermediate value theorem in Calculus is given in appendix C. Suppose now that we have $N$ Dirac delta centers along with the associated $N \times N$ matrix $\Phi$. As we remove one single center, its spectrum is determined from the principal submatrix of $\Phi$ of order $N-1$. From the Cauchy interlacing theorem, the eigenvalues of the matrix $\Phi$, with order $N$, are interlaced with those of any principal submatrix of $\Phi$ of order $N-1$. This means that the minimum eigenvalue of the principal submatrix of $\Phi$ is greater than or equal to the minimum eigenvalue of $\Phi$. Since the eigenvalues are increasing functions of $|E|$, we conclude that new minimum eigenvalue goes to zero at a lower point in the $|E|$ axis. Hence, the ground state energy increases for the new system. Actually, this argument can be applied to all other bound states as well. 
In other words, the bound state spectrum is shifted upwards as we remove one center from the system.

The analysis of the resonance phenomena  of the simpler version of the above model with positive strengths of the delta potential has been recently discussed by \cite{Sacchetti1}. Our formulation  here can be also useful for studying the resonances but we shall study the analysis of resonances for a future work

\section*{Acknowledgements}

The present work has been fully financed by TUBITAK from Turkey under the "2221 - Visiting  Scientist Fellowship Programme". We are very grateful to TUBITAK for this support. We also acknowledge Osman Teoman Turgut for clarifying discussions and his interest in the present research. Finally, this work
was also sponsored by the Spanish MINECO (MTM2014-57129-C2-1-P) and Junta de Castilla y Le\'on Project 
No. VA057U16. Finally, we thank the two anonymous reviewers whose comments improved this manuscript.

\section*{Appendix-A: The relation Between Our Formalism and the Self-Adjoint Extension Treatment of the Problem}
\label{AppAnew}

\setcounter{equation}{0}
\renewcommand{\theequation}{A-\arabic{equation}}

The resolvent formula associated with the finitely many point interactions is well known and given as  theorem 2.1.1 in \cite{Albeverio} 
\begin{eqnarray}
(-\Delta_{\alpha, Y} -k^2)^{-1}= G_k + \sum_{i,j=1}^{N} [ \Gamma_{\alpha, Y} (k)]^{-1}_{ij} (G_k(.,-a_j),.) G_k(.,-a_i) \;,
\end{eqnarray}
where $a_i \in Y$, $Y=\{ a_1, a_2, \cdots, a_N\}$, $\Delta_{\alpha, Y}$ is the self-adjoint extension of the free Hamiltonian defined in (2.1.5) in \cite{Albeverio}, $G_k=(-\Delta-k^2)^{-1}$, with $k^2$ is in the resolvent set of $-\Delta_{\alpha,Y}$, $\Im k>0$, and
\begin{eqnarray}
[\Gamma_{\alpha, Y}(k)]_{ij}=
\begin{cases}
\begin{split}
-\left[\alpha_{i}^{-1} +{i \over 2 k} \right]
\end{split}
& \textrm{if $i = j$}\;, \\ \\
\begin{split}
-  {i \over 2 k} e^{i k |a_i-a_j|}
\end{split}
& \textrm{if $i \neq j$}\;. \label{Albeverio1}
\end{cases}
\end{eqnarray}
Here the free resolvent kernel is given by $G_k (x-y) = {i \over 2 k} e^{i k |x-y|}$.
Theorem 2.1.3 in \cite{Albeverio} states that the bound state spectrum of the problem (as can be seen from the above resolvent formula) can be found from the zeros of the $\det [\Gamma_{\alpha, Y}(k)]$. It is important to notice that $\hbar=2m=1$ in \cite{Albeverio}.

Let us explicitly show that our heuristic formulation of the point interactions is completely consistent with the one given above. If we choose $\alpha_i \rightarrow - \lambda_i$ and restrict the matrix $\Gamma$ onto the negative real $E$ axis, namely $k=i \sqrt{|E|}$ ($E=k^2$ and $\Im k >0$), we obtain 
\begin{eqnarray}
[\Gamma_{\lambda, Y}(E)]_{ij}=
\begin{cases}
\begin{split}
\left[\lambda_{i}^{-1} -{1 \over 2 \sqrt{|E|}} \right]
\end{split}
& \textrm{if $i = j$}\;, \\ \\
\begin{split}
-  {1 \over 2 \sqrt{|E|}} e^{-\sqrt{|E|} |a_i-a_j|}
\end{split}
& \textrm{if $i \neq j$}\;. \label{Albeverio2}
\end{cases}
\end{eqnarray}
It is easy to show that our matrix $\Phi$ given in Eq. (14) is related by $\Gamma(E)$ through the following similarity transformation
\begin{equation}
S \; \Gamma(E) \; S^T = \Phi(E) \;,
\end{equation}
where $S= \diag(\sqrt{\lambda_1}, \cdots, \sqrt{\lambda_N})$. Here we have removed the subscripts $\lambda$ and $Y$ of $\Gamma$. This shows that $\det \Gamma(E) = \det \Phi(E)$. Therefore, our formulation is equivalent to the one given in \cite{Albeverio}. The only difference is that we exclude the case, where $\lambda_i$'s are infinite and consider only the bound states.

\section*{Appendix-B: A Proof of the Perron-Frobenius Theorem}
\label{AppA}

\setcounter{equation}{0}
\renewcommand{\theequation}{B-\arabic{equation}}

We give the simple proof of the Perron-Frobenius theorem for symmetric matrices given in \cite{Ninio}. Since the eigenvalues of $A$ are real and the sum of the eigenvalues are equal to the trace of $A$, we have $\tr A>0$. Then,  $\lambda>0$.
Let $(u_{j})$ be any real normalized eigenvector associated with the eigenvalue $\lambda$. Then, we have
\beqs
A u_{i}=\lambda u_{i}=\sum_{j}a_{ij} \; u_{j} \;, \label{eigenvalue problem perron one}
\eeqs
for $i=1,\ldots,n$ .  Setting $x_{j}=|u_{j}|$, we get
\beqs
0<\lambda=\sum_{ij}a_{ij} \, u_{i} \, u_{j}=\Big|\sum_{ij} a_{ij} \, u_{i} \, u_{j}\Big| \;,
\eeqs
and
\beqs
\lambda\leq\sum_{ij}|a_{ij}| \, |u_{i}| \, |u_{j}|=\sum_{ij}a_{ij} \, x_{i} \, x_{j} \;.
\eeqs
By means of the variational theorem, the right hand side is less than or equal to $\lambda$ (it is equal if and only if $(x_j)$ is the eigenvector  associated with the eigenvalue $\lambda$). Hence, we obtain
\beqs
\lambda \, x_{i}=\sum_{j}a_{ij} \, x_{j} \;, \label{eigenvalue problem perron two}
\eeqs
for $i=1,\ldots,n$. Therefore, if $x_{i}=0$ for some $i$, then because of $a_{ij}>0$ for all $j$, $x_{j}=0$ which cannot be true. Thus, $x_{j}>0$. This completes the first two part of the theorem.

For the third part, let us assume that $\lambda$ is degenerate. Hence, we can find two real orthonormal eigenvectors $(u_{j})$ and $(v_{j})$ associated  with $\lambda$. Suppose that $u_{i}<0$ for some $i$. From the addition of Eq. (\ref{eigenvalue problem perron one}) and (\ref{eigenvalue problem perron two}), we have
\beqs
\lambda(u_{i}+x_{i})=\sum_{j} a_{ij} \, (u_{j}+x_{j}) \Rightarrow \lambda \, (u_{i}+|u_{i}|)=\sum_{j}a_{ij} \, (u_{j}+|u_{j}|) \;.
\eeqs
Then, $u_{j}=-|u_{j}|$ for every $j$. If we assume that $u_{i}>0$ for some $i$ and subtracting Eq. (\ref{eigenvalue problem perron two}) from (\ref{eigenvalue problem perron one}), we obtain $u_{j}=|u_{j}|$. That means $u_{j}=\pm|u_{j}|$ and by applying the same procedure, we also have $v_{j}=\pm|v_{j}|$. Therefore,
\beqs
\sum_{j}v_{j} \, u_{j}=\pm \sum_{j}|v_{j} \, u_{j}| \;.
\eeqs
Since $|u_{j}|,|v_{j}|\neq0$ for all $j$, $|v_{j} \, u_{j}|\neq0$ which means that $u$ and $v$ cannot be orthogonal. Because of the contradiction with the first assumption, we conclude that $\lambda$ is non-degenerate.

As for the last part, let $(w_{j})$ be a normalized eigenvector associated with $\mu$ such that $\mu<\lambda$,
\beqs
\sum_{j}a_{ij} \, w_{j}=\mu \, w_{i} \;.
\eeqs
From the variational property and the non degeneracy of $\lambda$, we have
\beqs
\lambda >\sum_{ij}a_{ij} \, |w_{i}| \, |w_{j}|  \geq \Big| \sum_{ij}a_{ij} \, w_{i} \, w_{j} \Big|=|\mu| \;.
\eeqs

\section*{Appendix-C: A Proof of the Cauchy Interlacing Theorem}
\label{AppB}
 \setcounter{equation}{0}
\renewcommand{\theequation}{C-\arabic{equation}}

Here we give a simple proof of the Cauchy interlacing theorem using intermediate value theorem. This proof was originally given in \cite{cauchyinterlacing} and we give it here in order to be self-contained. 
Without loss of generality, the submatrix $B$ occupies rows $2,3, \ldots, N$ and columns $2,3, \ldots, N$. Then, the matrix $A$ has the following form:
\begin{eqnarray} A=
\left(
\begin{array}{cc}
 a & \mathbf{y}^{\dagger} \\
 \mathbf{y} & B
\end{array}
\right)
\end{eqnarray}
where $\dagger$ denotes the Hermitian conjugation. Since $B$ is also Hermitian, we can diagonalize it by a unitary transformation $U$:
\begin{eqnarray}
U^{\dagger} \, B \, U = D \;,
\end{eqnarray}
where $D=\diag(\mu_2, \mu_3, \ldots, \mu_N)$. For simplicity, let us define a new vector $\mathbf{z}=(z_2, z_3, \ldots, z_N)^{T} := U^{\dagger} \, \mathbf{y}$, where $T$ denotes the transposition. Here we only give the proof for the special case, where  $\mu_N < \mu_{N-1} < \cdots, < \mu_3 < \mu_2$ and $z_i \neq 0$ for all $i=2,3, \ldots, N$. The complete proof can be found in \cite{cauchyinterlacing}. Let
\begin{eqnarray} V=
\left(
\begin{array}{cc}
 1 & \mathbf{0}^{T} \\
 \mathbf{0} & U
\end{array}
\right) \;,
\end{eqnarray}
where $\mathbf{0}$ is the zero vector. Since $U$ is unitary, $V$ is also unitary. It is easy to see that
\begin{eqnarray} V^{\dagger} \, A \, V =
\left(
\begin{array}{cc}
 a & z^{\dagger} \\
 z & D
\end{array}
\right) \;.
\end{eqnarray}
Let us define the following function $f$:
\begin{eqnarray}
f(x):= \det(x I -A)\;,
\end{eqnarray}
where $I$ denotes the identity matrix. Since the determinant is invariant under unitary transformations, we have $f(x)=\det(x I - V^{\dagger} \, A \, V)$, or explicitly
\begin{eqnarray}
f(x)=
\left(
\begin{array}{cccccc}
 x-a & -z_2{}^* & -z_3{}^* & \cdots  & -z_{N-1}{}^* & -z_N{}^* \\
 -z_2 & x-\mu _2 & 0 & 0 & \cdots  & 0 \\
 -z_3 & 0 & x-\mu _3 & 0 & \ddots & \vdots  \\
 \vdots  & \vdots  & \vdots & \vdots & \ddots & 0 \\
 -z_{N-1} & 0 & \cdots  & 0 & x-\mu _{N-1} & 0 \\
 -z_N & 0 & 0 & 0 & 0 & x-\mu _N
\end{array}
\right) \;.
\end{eqnarray}
If we expand this determinant along the first row, we get
\begin{eqnarray}
f(x)=(x-a)(x-\mu_2) \cdots (x-\mu_N) - \sum_{i=2}^{N} f_i(x) \;,
\end{eqnarray}
where $f_i(x)=|z_i|^2 (x-\mu_2) \cdots \widehat{(x-\mu_i)} \cdots (x-\mu_N)$ for $i=2,3, \ldots, N$. Here the factor with a hat is deleted. Note that $f_i(\mu_j)=0$ for $j \neq i$ and
\begin{eqnarray}
f_i(\mu_i) 
\begin{cases}
\begin{split}
>0
\end{split}
& \textrm{if $i$ is even}\;, \\ \\
\begin{split}
<0
\end{split}
& \textrm{if $i$ is odd}\;.
\end{cases}
\end{eqnarray}
Since $f(\mu_i)= -f_i(\mu_i)$, the sign of $f(\mu_i)$ is opposite to that of $f_i(\mu_i)$. It is easy to see that $f(x)$ is a polynomial of degree $N$ with positive leading coefficient. Using this and the fact $f(x)$ is the characteristic equation for the matrix $A$,  and intermediate value theorem \cite{Calculus}, we conclude that there exist  $N$ roots $\lambda_1, \lambda_2, \cdots, \lambda_N$ of the equation $f(x)=0$ such that
\begin{eqnarray}
\lambda_N < \mu_{N} < \lambda_{N-1} < \mu_{N-1} < \cdots < \lambda_2 < \mu_2 < \lambda_1 \;. 
\end{eqnarray}
This is the result we wanted to show.


\begin{thebibliography}{99}

\bibitem{Demkov} Yu. N. Demkov, V. N. Ostrovskii,  \textit{Zero-range Potentials and Their
Applications in Atomic Physics} (Plenum Press, 1988).
%

\bibitem{Belloni Robinett} M. Belloni, R. W. Robinett, Phys. Rep. \textbf{540}, 25 (2014).

\bibitem{KPmodel} R. de L. Kronig, W. G. Penney, Proc. R. Soc. A \textbf{130}, 499 (1931).


\bibitem{CohenT} C. Cohen-Tannoudji,  B. Diu, F. Laloe F,
Quantum Mechanics, Vol. \textbf{1} (Wiley-Interscience, 2006).


\bibitem{L1} I. R. Lapidus, Am. J. Phys. {\bf 38}, 905 (1970).
%

%
\bibitem{Griffiths} D. J. Griffiths, \textit{Introduction to Quantum mechanics}
(Pearson Education, 2005).
%


\bibitem{LL} L. D. Landau, E. M. Lifshitz, \textit{Quantum Mechanics} (Pergamon Press, 1977).

%
\bibitem{Loudon} R. Loudon, Am. J. Phys. \textbf{27}, 649 (1959).
%



\bibitem{KJSQT} W. Kwong, J. L. Rosner, J. F. Schonfeld, C. Quigg, H. B. Thacker,
Am. J. Phys. \textbf{48}, 926 (1980).



\bibitem{Cohen} J. M. Cohen, B. Kuharetz, J. Math. Phys. \textbf{34}, 12 (1993).




\bibitem{Bhattacharyya} K. Bhattacharyya, R. K. Pathak, Int. J. Quantum. Chem. \textbf{59},
219 (1996).


\bibitem{Kar} S. Kar, R. R. Parwani, 
EPL \textbf{80}, 30004 (2007).


\bibitem{Vincenzo} S. De Vincenzo, Braz. J. Phys. \textbf{38(3a)}, 355 (2008).

\bibitem{Dutt} A. Dutt, T. Nath, S. Kar, R. Parwani, Eur. Phys. J. Plus \textbf{127}, 28 (2012). 



%
\bibitem{Meyer} C. D. Meyer, \textit{Matrix Analysis and Applied Linear Algebra} (SIAM, 2000). 
%
\bibitem{Albeverio} S. Albeverio, F. Gesztesy, R. Hoegh-Krohn, H. Holden, 
\textit{Solvable Models in Quantum Mechanics}, 2nd ed; (AMS, 2004).
%


\bibitem{BFV} G. Bonneau, J. Faraut, G. Valent,  Am. J. Phys. \textbf{69}, 3 (2001).
%
\bibitem{ACP} V. S. Araujo, F. A. B. Coutinho, J. F. Perez, Am. J. Phys. \textbf{72}, 2 (2004).
%




\bibitem{ASH}  R. B. Ash,  W. P. Novinger, {\it Complex Analysis} (Dover, 2007).
%
%

\bibitem{DenneryKrzywicki} P. Dennery, A. Krzywicki, \textit{Mathematics for Physicists} (Dover, 1996). 


\bibitem{Corless} R. M. Corless, G. H. Gonnet, D. E. G. Hare, D. J. Jeffrey, D. E. Knuth, Adv. Comput. Math. \textbf{5} 329 (1996).


\bibitem{SecilTunali} S. Tunal{\i}, Point interactions in quantum mechanics, Ms. Thesis, \.{I}zmir Institute of Technology, 2014.

\bibitem{Sacchetti} H. Kovarik, A. Sacchetti, J. Phys. A \textbf{43}, 155205 (2010).







\bibitem{Manoukian} E. B. Manoukian, \textit{Quantum Theory, A wide spectrum} (Springer, 2006).
%

\bibitem{feynman} R. P. Feynman, Phys. Rev. \textbf{56}, 340 (1939). 

\bibitem{hellmann} H. Hellmann, \textit{Einf\"{u}hrung in die Quantenchemie}  (Franz Deuticke, Leipzig, 1937), p. 285.



\bibitem{Vatsya} S. R. Vatsya, Phys. Rev. B \textbf{69}, 037102 (2004).
%
\bibitem{DJW} J. Dorsey,  C. R. Johnson, Z. Wei, Spec. Matrices \textbf{2}, 200 (2014).

\bibitem{BrownChurchill} J. W. Brown, R. V. Churchill, \textit{Complex Variables and Applications}, Eighth edition (McGraw-Hill, 2004).  


\bibitem{PalmaRaff} G. Palma, U. Raff, Can. J. Phys. \textbf{84}, 787 (2006).


\bibitem{reedsimonv4} M. Reed and B. Simon, 
\textit{Methods of Modern Mathematical Physics}, vol \textbf{IV},
(Academic Press, 1978).



\bibitem{ErmanTurgut} F. Erman, O. T. Turgut, J. Phys. A \textbf{43}, 335204  (2010).


\bibitem{Ninio} F. Ninio, J. Phys. A \textbf{9}, 1281 (1976).
%





\bibitem{Sacchetti1}  A. Sacchetti, J. Phys. A \textbf{49}, 175301 (2016).



\bibitem{cauchyinterlacing} S-G. Hwang, Am. Math. Mon.
\textbf{111}, 157 (2004).






%



\bibitem{Calculus} J. Steward, \textit{Calculus: Early Transcendentals}, 7th edition, (Brooks/Cole, Cengage Learning,  2012).






\end{thebibliography}
\end{document}